\documentclass[epj]{svjour}
\usepackage{graphicx}
\begin{document}
% 
%\title{Last Article Ever}
\title{Non-Affine Shear Modulus in Entangled Networks of Semiflexible Polymers}
\author{Hauke Hinsch \inst{1} \and Erwin Frey \inst{1}
}                     % Do not remove
\offprints{frey@lmu.de}          % Insert a name or remove this line
\institute{ Arnold Sommerfeld Center for Theoretical Physics and
  Center of NanoScience,\\
  Department of Physics, Ludwig-Maximilians-Universit\"at M\"unchen, \\
  Theresienstrasse 37, D-80333 M\"unchen, Germany,}
\date{Received: date / Revised version: date}
\abstract{We investigate the viscoelastic properties of entangled
  networks of semiflexible polymers. At intermediate time scales the
  elastic response of these networks to shear deformation is described
  by the plateau modulus $G$. Different scaling laws with polymer
  concentration $c$ have been proposed based on the assumption that
  the deformation field is affine on all length scales. We develop a
  numerical approach that allows to calculate the modulus via free
  energy changes for both affine and non-affine deformations. The
  non-affine deformation field is obtained by a free energy
  minimization. Our findings allow for a confirmation of a power law
  $G \propto c^{7/5} l_p^{-1/5}$ with polymer concentration $c$ and
  persistence length $l_p$ and furthermore quantify the systematic
  deviations due to the affinity assumption.}

% end of PACS codes
%
\maketitle

\section{Introduction}

Semiflexible polymers have always been an interesting testing ground
for concepts of statistical physics, since excitable energies are on
the order of $k_\mathrm{B} T$, fluctuations are important and entropic
effects compete with bending energy. However, with a few exception
like the tangent-tangent correlation function
\cite{doi86,rubinstein03}, moments of the end-to-end distance
\cite{saito67} in the wormlike-chain model \cite{kratky49,saito67} or
the end-to-end distribution function \cite{wilhelm96} in the
weakly-bending rod approximation, a multitude of interesting single
polymer properties are not accesible to analytic solutions. This
applies even more so to many-polymer systems like networks or
solutions. In these cases, inter-polymer interactions or topological
constraints complicate the theoretical treatment and the mere number
of constituents and degrees of freedom clearly renders any analytical
attempt to calculate the partion sum unfeasible. Our line of approach
is therefore to reduce the many-body problem to a single-polymer
description. A prominent example of such an approach is the tube model
introduced by de Gennes \cite{gennes79} and Doi and Edwards
\cite{doi86} that describes the combined effect of neighboring
polymers by a mean-field potential. These theoretical models for
single polymers at hand, one can proceed to describe macroscopic
properties of a polymeric material. Here it is essential to build on
the microscopic constituents without loosing emerging collective
properties of the macroscopic material.

These material properties of polymer networks are of a stunning
variety and complexity. Spurred by the biological importance of
networks of filamentous actin (F-actin), intensive experimental
research has recently focussed on this model system. In the presence
of cross-links polymer solution form permanent gels or bundles with
different elastic properties determined by a competition of bending
and stretching modes \cite{mackintosh95,shin04,wilhelm03,head03}.
Recently, also the importance of non-affine deformations has been
pointed out \cite{heussinger06}. For purely entangled solutions a
strong dependence of both the storage and the loss modulus on
frequency was observed \cite{amblard96,gittes97,morse01,gardel03}.
Furthermore, similarities to glassy systems were reported
\cite{kroy07,semmrich07}.  Upon application of larger stress a
non-linear regime was investigated and shear stiffening of the network
was observed \cite{xu00,semmrich08}.  Theoretical analysis identified
different scaling regimes for the moduli with frequency
\cite{morse98}.  Regarding the scaling of the plateau modulus with
concentration Isambert and Maggs derived a power-law of $7/5$ from a
simple scaling argument \cite{isambert96} based on the deformation of
confinement tubes.  Other theories attribute the material response to
the suppression of undulations by stretching
\cite{mackintosh95,kroy96}.  They predict a scaling with a
considerably larger exponent, but seem to disagree with experimental
data \cite{hinner98,xu98,gardel04}.  Recent experimental work
\cite{tassieri08} claims to have verified an exponent of $4/3$
predicted by an elastic medium theory \cite{morse01}. Common to all
these approaches is the assumption that the macroscopic deformation
field is assumed to be affinely transmitted to all length scales. We
will present an approach that permits to go beyond this assumption and
investigate resulting differences.

We proceed as follows: in Section \ref{sec:system} we define the
system under consideration. We introduce the Hamiltonian of the
network, review the simplifications that lead to the tube model and
prior work on the plateau modulus, and emphasize the mean-field nature
of the tube model and the assumption of affine displacement. In
Section \ref{sec:numeric} we present our approach to numerically
compute the free energy of the system by a reduction to two
dimensions. We explain the effect of global shear on the microscopic
constituents of the network and introduce a free energy minimization
procedure that results in a non-affine deformation field. We proceed
in Section \ref{sec:results} with the presentation and interpretation
of our results before we conclude in Section \ref{sec:summary}.

\section{System Definition} \label{sec:system}

We consider a network of semiflexible polymers that only interact via
a hard-core potential and thus constrain each other topologically. The
polymer density is given by the number $\nu$ of polymers of length $L$
per unit volume. The stiffness $\kappa$ of the polymers gives rise to
a persistence length of $l_p=\kappa / k_\mathrm{B} T$. The
configuration of the $i$-th polymer in space, ${\bf r}_i(s)$, is
parameterized by arc length $s$ and the average distance between the
networks constituents is characterized by the mesh size,
$\xi=\sqrt{3/\nu L}$. Concerning the mechanical response of the
network, we are interested in the dynamic processes that occur after a
deformation has driven the system out of equilibrium. These relaxation
processes occur on different time scales. While generally every stress
can relax by reptation of the polymers, this process is dramatically
slowed down due to topological constraints in crowded environments
\cite{hoefling08b,hoefling08c}. On intermediate time scales relevant for the
plateau modulus it can be assumed that the constraints imposed by
surrounding polymers can not be overcome and that the center of mass
of all filaments does not change substantially. A given polymer $i$ is
then described by the worm-like chain model \cite{kratky49,saito67}
with a Hamiltonian that has contributions from intra-polymer bending
and from interactions with the neighboring filaments in the solution.
This can be written as
\begin{equation} \label{eq:hamilton1} H_i=\frac{\kappa}{2} \int_0^L ds
  \left( \frac{\partial^2 {\bf r}_i(s)}{\partial s^2} \right)^2 +
  \sum_{j \neq i}
  \Theta_{i,j}^\mathcal{I} \;,
\end{equation}  
where the function $\Theta_{i,j}^\mathcal{I}$ formally describes the
hard-core interaction between the polymers $i$ and $j$ under the
initial topology $\mathcal{I}$. Note that this function does not only
depend on the configurations ${\bf r}_i(s)$ and ${\bf r}_j(s)$ of the
polymers like a conventional hard-core potential that would always be
zero in a system of mathematical lines without excluded volume.
Instead it crucially depends on the inital topology and rules out
polymer crossing by returning an infinite energy if the two polymers
interpenetrate \footnote{For times far larger than the relevant time
  scale for the plateau modulus the contributions
  $\Theta_{i,j}^\mathcal{I}$ vanish as topological contraints can be
  overcome by reptation. This signifies that long time averages will
  reproduce the results obtained from an ensemble average with respect
  to the free polymer Hamiltonian.}. The Hamiltonian of the complete
system is obtained as $H = \sum_i H_i$.

Given the form of the Hamiltonian (\ref{eq:hamilton1}) a calculation
of the free energy $F$ from the partition sum $Z$ as $F=-k_\mathrm{B}
T \ln Z$ is obviously not feasible as the partition sum 
\begin{equation} \label{eq:Z1}
Z = \Pi_i \int {\cal D}[{\bf r}_i(s)] \exp[-H/(k_\mathrm{B}T)]
\end{equation}
amounts to multiple path integrals of a highly convoluted integrand.

A simplifying description of the system was proposed with the famous
tube model \cite{gennes79,doi86} where the combined effect of the
fluctuating neighbor polymers on a single test polymer is described by
an effective harmonic potential. While the tube model is the
foundation for theories for different properties of semiflexible
polymer networks like tube diameter \cite{semenov86,morse01,hinsch07}
or viscoelasticity \cite{isambert96}, it has to be kept in mind that it
only provides a mean-field description of the microscopic
constituents. In the remainder of this section we will review how free
energies and mechanical properties can be derived from the tube model
and point out possible shortcomings of this coarse-grained frame of
description.

The tube model can be applied to solutions of semiflexible polymers
were the confinement of a single polymer by its neighbors is
sufficiently strong to guarantee that the transversal undulations of
the polymer do not deviate far from an average contour in space - the
tube backbone ${\bf r}^0(s)$. This is the case, if persistence length
and polymer contour length are substantially larger than the typical
void spaces in the mesh of surrounding polymers, thus $L, l_p \gg \xi$
as e.g. given for most F-actin networks. The complex sum in the second
term of the Hamiltonian (\ref{eq:hamilton1}) can then conveniently be
substituted by a harmonic potential with average strength $\gamma$ and
minimum at the tube backbone:
\begin{equation} \label{eq:hamilton2} H(\gamma,\kappa)=\int_0^L ds
  \left[\frac{\kappa}{2} \left( \frac{\partial^2 {\bf r}(s)}{\partial
        s^2} \right)^2+\frac{\gamma}{2} \left( {\bf r}(s)-{\bf r_0}(s)
    \right)^2 \right] \;.
\end{equation} 
As pointed out by Odijk \cite{odijk83} it is instructive to introduce
an additional length scale $L_d \approx (\nu L)^{-2/5} l_p^{1/5}$
known as deflection or Odijk length. While the length scales $L$ and
$l_p$ describe the properties of the single polymer and the length
scale $\xi$ describes the network, the deflection length $L_d$
captures the interaction between both. It can be interpreted as a
measure for the distance between two collisions between the encaged
polymer and the tube walls and therefore the number of collisions of a
polymer is given as $L/L_d$.  This is also reflected in the free
energy cost $\Delta F$ that arises from the restriction of the test
polymer to a tube and is obtained by a path integration of
(\ref{eq:hamilton2}) over all polymer configurations
\cite{burkhardt95} as
\begin{equation} \label{eq:free_energy}
\Delta F=\sqrt{2} k_\mathrm{B} T \frac{L}{L_d} \;.
\end{equation}
This signifies that at a spacing of $L_d$ between two collisions every
of the $L/L_d$ contact points between polymer and tube contributes one
$k_\mathrm{B} T$ to the confinement free energy.

Having derived the free energy in the coarse-grained tube model, the
next step is to analyze the change in free energy at mechanical
deformation to obtain the plateau modulus. As the polymers are
described in terms of their tubes, it is obvious to investigate the
effect of deformation on the tubes for which the free energy is known
as reasoned above. Together with a scaling law $d \propto c^{-3/5}
l_p^{-1/5} $ for the tube diameter $d$ derived by Semenov
\cite{semenov86} this line of reasoning was first used by Isambert and
Maggs \cite{isambert96} to establish a scaling relation between
plateau modulus and concentration. They argue that the macroscopic
shear deformation is affinely passed down to the tubes that are
compressed or stretched depending on their orientation to the shear.
The resulting change of the tube diameter causes a change in the
deflection length $L_d$ and with the help of (\ref{eq:free_energy})
the resulting modulus scales as
\begin{equation} \label{eq:modul1}
G \propto c^{7/5} l_p^{-1/5}\;.
\end{equation} 
The same scaling was also obtained by other descriptions, {\it e.g.}
by a modified Onsager theory for confinement tubes \cite{hinner98}.
In an endevour to arrive at a quantitative theory for the plateau
modulus Morse \cite{morse01} has proposed two conceptually different
approaches: a detailed microscopic description of the topological
constraint imposed by neighboring polymers leads him to the prediction
of a modulus scaling with $c^{7/5}$ and a quite different approach
yields a scaling of $G \propto c^{4/3}$ derived by a self-consistent
treatment of the network as an elastic continuum.  Since these values
are numerically quite close, a decisive distinction between the models
has not yet been possible with the accuracy of available experimental
data.

It has to be kept in mind that all these theoretical approaches are
implicitly build on the assumption that the tube contour deforms
affinely with the macroscopic strain. This is evidently only a very
coarse-grained description of the system's response. While stress
relaxation by slow processes like reptation is obviously not relevant
on the time scale of the plateau modulus, it is however possible that
faster relaxation processes cause a tube contour that differs from the
contour obtained by affine displacement. Our goal is to implement this
relaxation processes by a free energy minimization in a numerical
solution of the plateau modulus and investigate the quantitative and
qualitative differences to the affine model. The detailed setup of
this approach is discussed in the following section.

\section{Numerical Solution} \label{sec:numeric}

Our approach is to obtain a numerical solution of the partition sum
(\ref{eq:Z1}) for a test polymer in a typical network of semiflexible
polymers by averaging over all allowed configurations of neighboring
polymers. The advantage of this approach is a microscopic description
of the Hamiltonian. In contrast to the tube-model that only provides a
coarse-grained description of the surrounding polymers, it accounts
for the detailed interactions in a given realization of disorder. This
permits to investigate the effect of local non-affine deformations of
the encaged test polymer. Since the distribution of obstacle polymers
around a given test filament is quite heterogeneous, it is expected
that these non-affine deformations result in a lower global free
energy. Our aim is to find this free energy minimum by a numeric
minimization procedure.

\subsection{Reduction to 2D}
\begin{figure}[htbp]
\center
\includegraphics[width=\columnwidth]{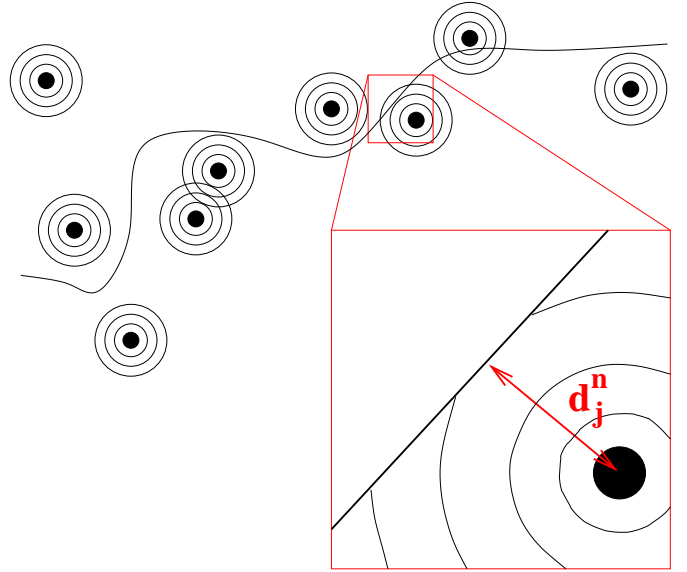}
\caption{Fixed polymer in an array of fluctuating point obstacles
  (black points).  The interaction between the polymer and the
  obstacles as the fluctuations of the $j$-th point obstacle are
  hindered by the polymer at a distance $d^n_j$ (see inset).}
\label{fig:fixed_poly}
\end{figure}

To reach this goal, we start by decomposing the transverse undulations
of the test polymer into two independent components as previously
described \cite{hinsch09}. For one component the Hamiltonian thus
simplifies to the description of a two-dimensional polymer in a plane
surrounded by a certain number $N_\mathrm{obs}$ fluctuating point-like
obstacles (see Fig.  \ref{fig:fixed_poly}).
The point-like obstacles are subjected to a harmonic potential with
strength $\gamma$ around an equilibrium position ${\bf p}_j^0$ with
$j=1,..,N_\mathrm{obs}$. The parameters $N_\mathrm{obs}$ and $\gamma$
can be chosen to self-consistently represent a network of a specific
concentration \cite{hinsch07}.  Therefore the system is completely
described by the two-dimensional contour ${\bf r}(s)$ of the test
polymer and $N_\mathrm{obs}$ two-dimensional vectors ${\bf p_j}$
describing the positions of the obstacles in the plane. The
Hamiltonian thus reads $H=H^p + \sum_{j=1}^{N_\mathrm{obs}}
H^\mathrm{obs}_j$ where $H^p$ is the bending energy contribution from
the polymer
\begin{equation}
  H^p=\frac{\kappa}{2} \int_0^L ds
  \left( \frac{\partial^2 {\bf r}(s)}{\partial s^2} \right)^2
\end{equation}
and the $H^\mathrm{obs}_j$ are the contributions from the obstacle
points
\begin{equation} \label{eq:hamilton3}
  H^\mathrm{obs}_j=\frac{\gamma}{2} \left( {\bf p}_j-{\bf p}^0_j
  \right)^2 + \Theta[{\bf r}(s),{\bf p}_j,{\bf p}^0_j] \;,
\end{equation}
where the topological constraint of uncrossability is again described
by a function $\Theta$ that returns an infinite energy if the polymer
and a point obstacle cross. In contrast to the three-dimensional case
Eq.~(\ref{eq:hamilton1}) this function can now easily be grasped
geometrically as we will show below. In calculating the corresponding
partition sum we have to solve
\begin{equation}
  Z=\int {\cal D}[{\bf r}(s)] \int \prod_{j=1}^{N_\mathrm{obs}} d{\bf p}_j \exp[-\beta H^p] 
    \exp[-\beta \sum_{j=1}^{N_\mathrm{obs}} H^\mathrm{obs}_j] \;.
\end{equation}
While the description has now reduced to a single path integral, an
analytical solution is still complicated by the topological
constraints. However, for a specific polymer contour, the integration
over the degrees of freedom of the obstacles is straightforward and
can easily be carried out analytically as the only topological
restriction for each point obstacle is posed by the test polymer.
Since the polymer is mostly straight on the length scale of the
typical fluctuation width of an obstacle, we assume that integration
over the obstacle potential is performed only in the half-space that
is limited by the test polymer at a nearest normal distance $d^n_j$ as
depicted in Fig.~\ref{fig:fixed_poly}. The partition sum is then
written as:
\begin{equation} \label{eq:z2}
Z=\int {\cal D}[{\bf r}(s)] \exp[-\beta H^p]
\prod_{j=1}^{N_\mathrm{obs}} \frac{\pi}{\gamma} \mathrm{erfc} \left[-d^n_j \sqrt{\frac{\gamma}{2}} \right]  \;.
\end{equation}
To solve the remaining path integration we chose to apply a saddle
point approximation in which we first assume the test polymer to be
immobile. Then we find the fixed contour that maximizes the partition
sum and thereby minimizes the free energy and finally add fluctuations
around this minimum. In the first step we are faced with the
minimization problem depicted in Fig.  \ref{fig:fixed_poly}.  An
immobile polymer with associated bending stiffness is placed in an
array of fluctuating obstacles. The free energy is composed of the
bending energy of the polymer and the entropic contributions from the
obstacles. It is obtained from the partition sum from Eq.
(\ref{eq:z2}) and is a function of the polymer contour ${\bf r}(s)$
alone:
\begin{equation} \label{eq:f1}
  F({\bf r}(s))=H^p({\bf r}(s)) +
  \sum_{j=1}^{N_\mathrm{obs}} - k_\mathrm{B}
    T \ln \left(\frac{\pi}{\gamma} \mathrm{erfc}\left[-d^n_j
    \sqrt{\frac{\gamma}{2}}\right] \right) \;.
\end{equation}
We are now looking for the contour ${\bf r^0}(s)$ of the polymer that
minimizes the free energy for a given initial setup of obstacles.

\subsection{Free Energy Minimization}
Technically, this contour is obtained as follows. We start from a
given initial polymer configuration and choose $N_{min}$ 
nodes as a discretization along its contour. This reduces the required
minimization to $N_{min}$ dimensions and permits to obtain a feasible
computation time by a suitable choice of
$N_{min}$. To find the minimum of the free energy, Eq.÷(\ref{eq:f1}), we 
move these nodes certain distances transverse to the 
present polymer contour, and define the new trial contour as a
cubic spline through the new node positions; here one has to make sure that
total contour length is kept constant. With this new contour both the bending
energy and the entropic contribution to the free energy is computed according 
to Eq.÷(\ref{eq:f1}). These steps of moves transverse to the immediately preceding contour are repeated until the minimum of the free energy is reached. The minimization algorithm we used is based on the AMOEBA
\cite{nr} implementation of the Nealder-Mead method \cite{neldermead}.

\subsection{Mode Representation}
\begin{figure}[htbp]
\center
\includegraphics[width=\columnwidth]{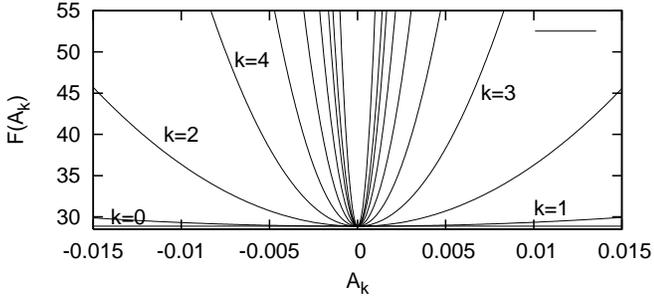}
\caption{Increase of free energy with amplitude for different modes in
  a given random array of obstacles. }
\label{fig:fluc_poly2}
\end{figure}

Now that we have found the polymer contour of lowest free energy, we
proceed to add the transversal fluctuations of the polymer in form
bending modes.
\begin{figure}[htbp]
\center
\includegraphics[width=\columnwidth]{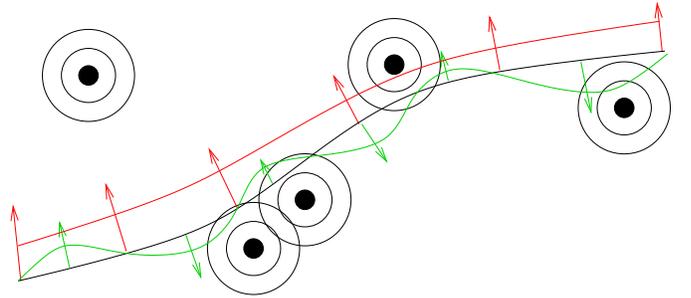}
\caption{Modes $k=0$ (red) and $k=3$ (green) around the contour of
  minimal free energy (black).}
\label{fig:fluc_poly}
\end{figure}
The contour of minimal free energy ${\bf r^0}(s)$ can
be interpreted as the backbone of the test polymer's confinement tube
or the contour with the highest probability. All deviations from this
contour have higher free energy and thus a smaller probability. We
model the thermal undulations of the polymer around this tube backbone
by cosine modes $u_k(s)$ in the form 
\begin{equation} \label{eq:eigenmode} r_\perp(s)= \sum_k u_k(s) =
  \sum_k A_k \cos(\frac{s k}{L})
\end{equation}
where $k$ is the mode number and $A_k$ is the mode amplitude. In this
representation the mode $k=0$ is simply a transversal displacement of
every point of the polymer normally to ${\bf r^0}$. A visualization of
this mode and the mode $u_{k=3}$ is depicted in Fig.
\ref{fig:fluc_poly}. For a specific mode we can monitor the resulting
free energy as a function of the mode amplitude $A_k$ as exemplary
shown in Fig. \ref{fig:fluc_poly2}. The result is a harmonic function
$F(A_k)=\omega_k/2 A_k^2$ where $\omega_k$ can be determined from the
plots for every mode.
The Hamiltonian for a certain contour that is in the representation
(\ref{eq:eigenmode}) fully characterized by the set of coefficients
$\{A_k\}$ is then given as
\begin{equation}
H(\{A_k\})=\frac{1}{2} \sum_k \omega_k A_k^2
\end{equation}
The desired partition function can thus easily be obtained by
integration over the mode amplitudes
\begin{equation} \label{eq:z_modes}
Z=\prod_k dA_k \exp[-\beta \omega_k A_k^2 / 2]=\prod_k \sqrt{\frac{2
    \pi}{\beta \omega_k}} \;.
\end{equation}
The free energy cost of confinement is obtained by calculating for every 
mode the free energy difference between a confined and a free polymer.
Since $\omega_k \propto k^4$ (compare Fig.
\ref{fig:fluc_poly2}) this difference becomes small quickly with increasing mode number $k$
such that it suffices to compute the free energy difference for the first few
modes.

\subsection{Shear Deformation}

Having developed an approach to calculate the free energy of a
fluctuating polymer in an array of fluctuating topological
constraints, we can proceed to investigate the free energy change as a
reaction to shear deformations. A simple example of such a deformation
is a global shear deformation of a macroscopic sample. If the sample
is an equilibrated network of semiflexible polymers, the result of the
shear deformation will be a rise in free energy and consequently a
force counter-acting the deformation. The system is thus perturbed by
the deformation and brought to a non-equilibrium state which will
immediately be followed by relaxation processes. These relaxation
processes occur on very different time scales for the different length
scales in the network. This is the reason for the frequency dependence
of the modulus. 

\begin{figure}[htbp]
\center
\includegraphics[width=\columnwidth]{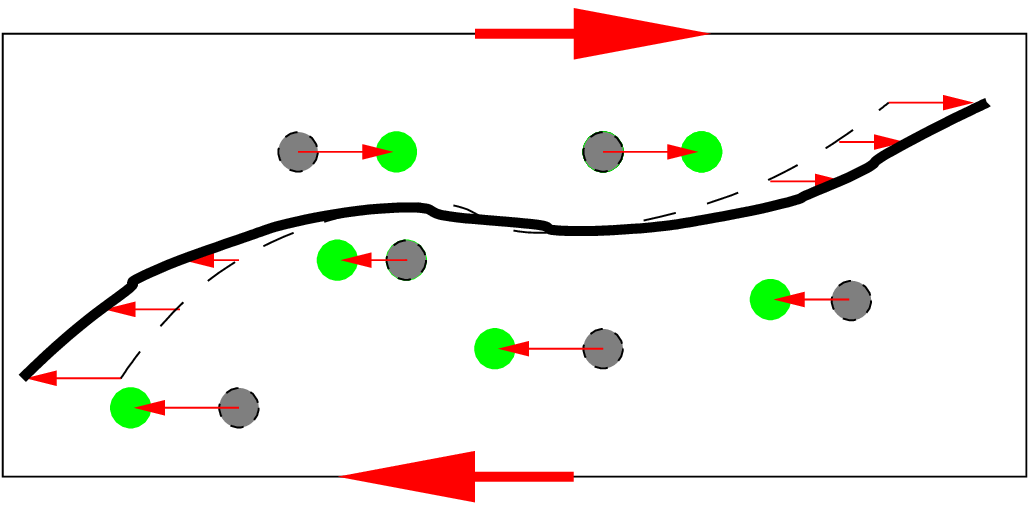}
\includegraphics[width=\columnwidth]{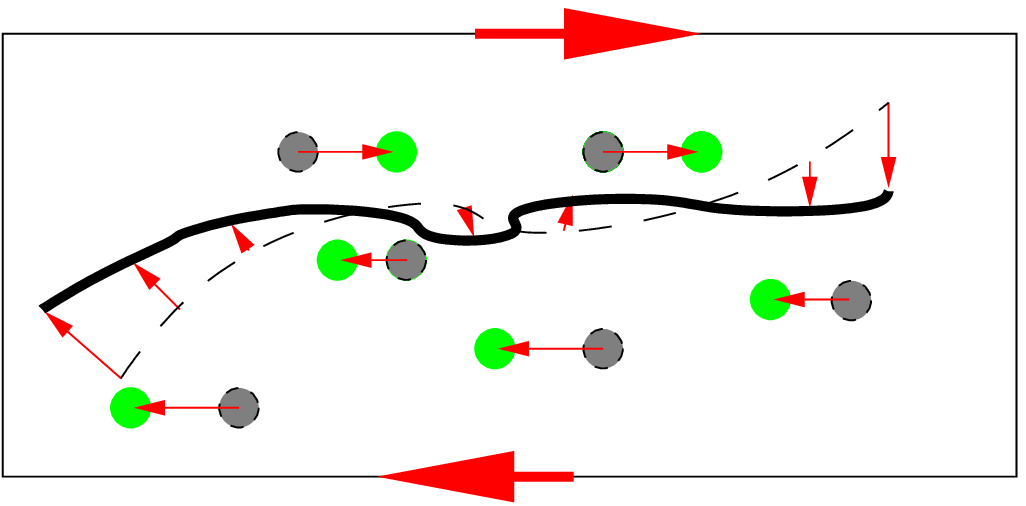}
\includegraphics[width=\columnwidth]{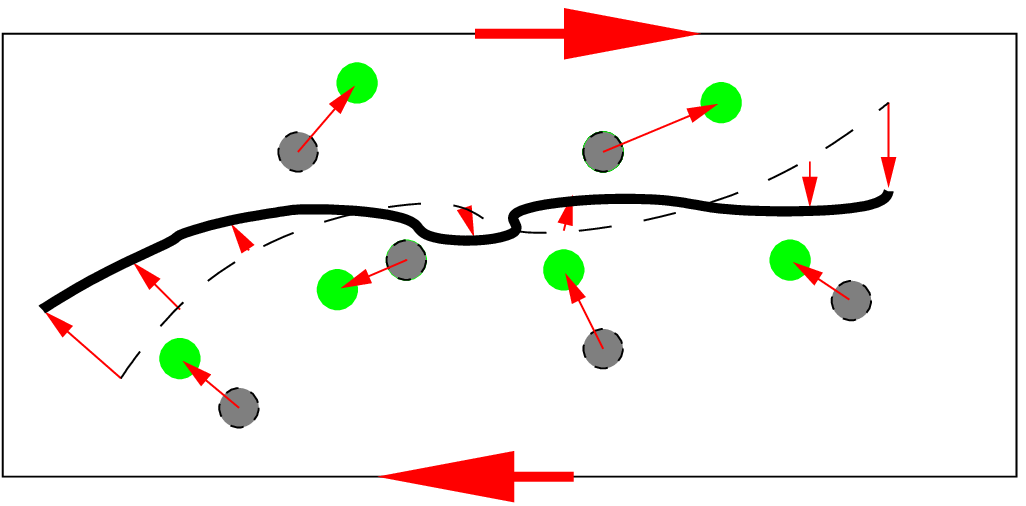}
\caption{{Different levels of affinity in shear deformations.  ({\bf
      top}) Obstacle fluctuation centers and tube backbone follow the
    macroscopic shear deformation (large red arrow) affinely. ({\bf
      middle}) While the obstacle fluctuation centers follow the
    macroscopic shear deformation affinely, the tube backbone can
    deviate from the affine deformation field in order to minimize the
    global free energy. ({\bf bottom}) Also the obstacle point can
    react by non-affine deformation to the macroscopic shear.}}
\label{fig:shear}
\end{figure}

For very long time for instance, the network is able to completely
relax the deformation stress by reptation thereby recovering the
equilibrium value of free energy and resulting in a vanishing modulus.
We can assume that at the time scale of the plateau modulus the
encaged polymers have completely experienced their immediate
surroundings but no large scale relaxation by network rearrangement
has occurred. Thus in measuring the plateau modulus the tube has
sufficient time to form before the deformation field changes again.
This argument is the foundation for the assumption of affine
displacement of the tube's contour and size. As it is unknown how
exactly the macroscopic stress is passed on to the microscopic
constituents of the network, it is commonly assumed that the
deformation field follows the macroscopic stress on all length scales.
Since at the time scale of the plateau modulus the tube is the
relevant quantity, the tube centers or backbones are displaced
affinely with the global shear.  Translated to our two-dimensional
plane of observation, this signifies that the tube contour ${\bf
  r^0}(s)$ and the centers of the obstacle tube $p^0_j$ are deformed
affinely as depicted in Fig. \ref{fig:shear} ({\it top}).  The free
energy of this new configuration can be calculated as shown above. The
modulus $G_A$ to be determined from this resulting free energy change
should be equivalent to the modulus obtained from any theoretical
treatment that is based on the assumption of affine displacement.

If we take a closer look at the processes at tube formation, it
becomes clear that the tube obtained by affine deformation is not
necessarily a valid description of the physical reality. To this end
we relax the assumption of affine displacement in such a way, that we
now only let the obstacle fluctuation center $p^0_j$ deform affinely
as illustrated in Fig. \ref{fig:shear} ({\it center}). If this new
configuration of the obstacle array is taken for granted, we have to
ask the question how the test polymer encaged by the obstacles reacts
to this conformational change. Out of all possible deformations of the
test polymer only the one with the lowest free energy will actually be
realized. Obviously, this new tube contour will only in very few cases
equal the tube contour obtained by affine displacement of the original
contour. The resulting free energy of this deformation will thus be
less or equal to the free energy obtained by affine deformation and we
can therefore also state that the resulting modulus $G_{NA} \leq G_A$.
Technically, the free energy difference of this non-affine deformation
of the tube contour is obtained by displacing the $p^0_j$ affinely
with the macroscopic stress (see \ref{sec:appendix1}) and then
applying again the free energy minimization as explained above to find
the new tube contour.

Of course, also the fact that we deform the obstacle points affinely
implies an assumption. In the actual physical system the obstacle
points and with it the tube centers of the neighboring polymers are
free to change their position in order to reach a global state of
lower free energy (see Fig. \ref{fig:shear} ({\it bottom}). As these
neighboring polymers however couple to other polymers outside our
plane of observation, the incorporation of this feature would be
tantamount to a minimization in all degrees of freedom of the network.
This is obviously out of range of a numerical solution. The resulting
modulus of a complete free energy minimization is the modulus $G$ that
would be observed in experiments. The modulus determined by our
approach constitutes an upper bound for the experimental values and
thus $G \leq G_{NA}$.

\section{Results} \label{sec:results}

In the previous section we presented an approximative numerical
solution to the problem of finding the free energy change under shear
deformation of a single probe polymer. This is obviously a quantity
that can not be observed experimentally, but it can serve as the basis
for the calculation of the macroscopic plateau modulus. To obtain this
observable we have to add up the contributions from all polymers in
the network under consideration. One single specific polymer is
described in terms of the two dimensional plane of observation whose
orientation in space is described by a set of angles
$(\theta,\phi,\psi)$ as described in Appendix \ref{sec:appendix1}.
Consequently, we have to perform an average over these isotropically
distributed angles and furthermore we have to average over the
quenched disorder that is generated by the different configurations of
point-like obstacles in the observation plane. With the shear
parameter $\Gamma$ this procedure finally returns an average free
energy function $\Delta F(\Gamma)= g \Gamma^2/2$ from which the
macroscopic modulus is obtained as $ G = 2 \nu g$. The factor $2$
stems from the fact that every polymer is described by two planes of
observation corresponding to the two components of transverse
fluctuation.

\begin{figure}[htbp]
  \center
  \includegraphics[width=\columnwidth]{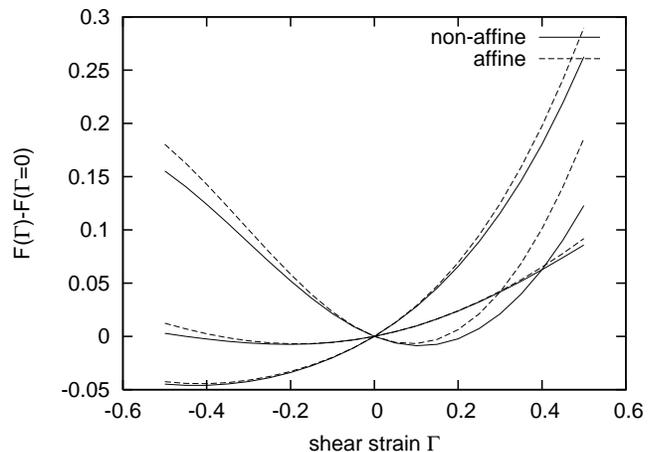}
\caption{Free energy change with shear $\Gamma$ for three different
  realizations of a test polymer in a network.}
\label{fig:snapshots}
\end{figure}

For a single polymer in a plane and one specific realization of
obstacle disorder the resulting free energy function is exemplary
shown in Fig. \ref{fig:snapshots} {\it (top)}. 
Since the absolute value of the free energy differs strongly with the
actual obstacle configuration all plots have been rescaled to the free
energy value at $\Gamma=0$. Obviously, for a single polymer in
different specific realizations of obstacle disorder and different
orientations of the plane of observation to the applied shear the
resulting form of the free energy is highly variable. Furthermore, the
free energy minimum is in general not at the point of zero shear. This
feature however, should of course be fulfilled for the free energy
function that is obtained by summing up all constituents in an
macroscopic sample at equilibrium. We chose to use this requirement as
a verification for an sufficient sampling over disorder. The location
of the accumulated free energy minimum initially strongly oscillates
with the number of samples but finally converges to $\Gamma=0$. For
every data point we average over a sufficient number of disorder
samples until this criterion is fulfilled.

\begin{figure}[htbp]
\center
\includegraphics[width=\columnwidth]{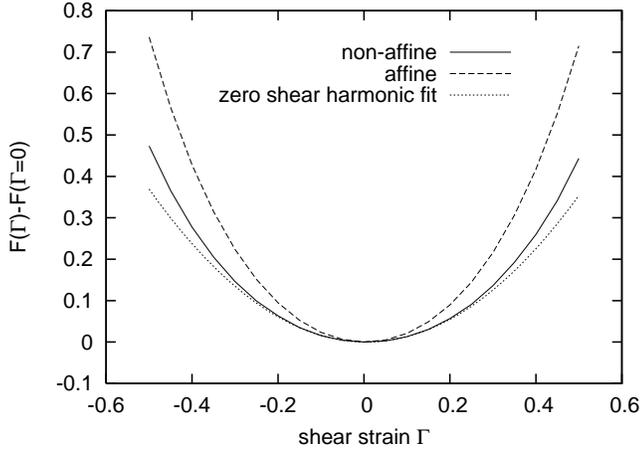}
\caption{Free energy change obtained by averaging over quenched
  disorder and orientation where the lower free energy curve is due to
  energy minimized non-affine deformations. Harmonic fits are only a
  valid approximation in the linear regime at small shears.}
\label{fig:combined}
\end{figure}

An example of the resulting averaged free energy function is shown in
Fig. \ref{fig:combined}. 
As expected the minimum is at zero shear where the system is at
equilibrium. At the application of small shear the free energy rises
in a harmonic fashion which would entail a linear restoring force in
an experimental measurement. Furthermore, it can be seen that the free
energy obtained in the affine approximation is always above the
non-affine free energy that was obtained by the minimization procedure
explained above. We determine the modulus by an harmonic fit at small
shear strains. At higher strains however, the free energy function is
no longer faithfully described by this fit, but features a stronger
slope. This signifies the onset of non-linear forces.

\subsection{Affine vs. Non-Affine}

\begin{figure}[htbp]
  \center
  \includegraphics[width=\columnwidth]{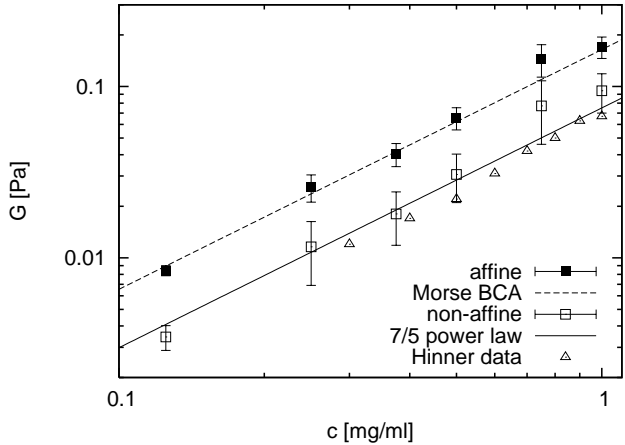}
\caption{Moduli
  resulting from affine and non-affine displacement of the tube
  contour as a function of actin concentration comply with a $7/5$
  power law. The affine modulus (filled squares) is in the range of
  the prediction by Morse \cite{morse01} while the non-affine modulus
  (open squares) is a factor two to three lower but slightly above the
  experimental measurements (open triangles) by Hinner
  \cite{hinner98}.}
\label{fig:gc}
\end{figure}

We determined the resulting plateau moduli as a function of different
system parameters. Fig. \ref{fig:gc} illustrates the scaling of the
modulus with polymer concentration $c$. 
Both the affine and the non-affine modulus show good agreement to a
$7/5$ power law with concentration. The non-affine modulus is
considerably below the affine modulus. This confirms the initial
assumption that a deformation field that assumes affine displacement
on all length scales indeed over estimates the system's response. The
non-affine deformation that is obtained by permitting the encaged
polymer to find its tube of minimal free energy leads to a lower
modulus. Comparing the moduli from our affine calculation with the
prediction for the absolute plateau modulus by Morse's ``Binary
Collision Approximation'' \cite{morse01}, shows sound consistency. In
the realm of the restriction of affine displacement, our work can be
seen as a numerical confirmation.  However, the moduli obtained by
experimental measurements \cite{hinner98} are considerably lower and
prove that the physical reality is closer to a non-affine deformation
field. Of course, the detailed nature of this field is not accessible
to experiments but our data suggests that the proposed model of an
affine displacement of neighbors combined with an non-affine
displacement of the tube is an appropriate approximation. The
non-affine moduli obtained by this approach only show slight
overestimations of the experimental results and it can be argued that
this is due to possible additional non-affinities in the obstacle
displacement.

\subsection{Scaling with Persistence}

\begin{figure}[htbp]
\center
\includegraphics[width=\columnwidth]{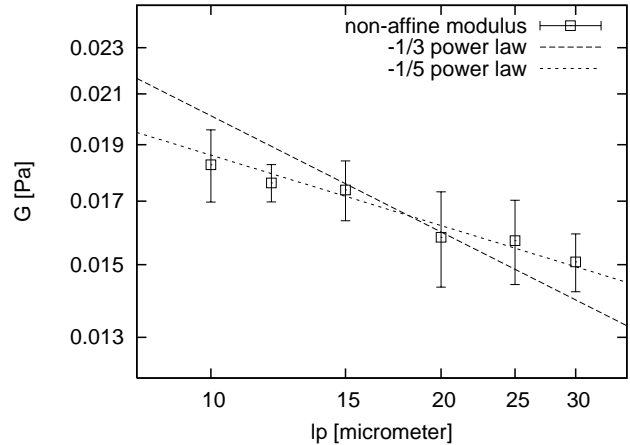}
\caption{The non-affine plateau modulus clearly shows a $-1/5$ power
  law dependence on persistence length.}
\label{fig:glp}
\end{figure}

Finally, we determined the scaling of the plateau modulus with
persistence length $l_p$. The decrease of the modulus with increasing
polymer stiffness is depicted in Fig. \ref{fig:glp} and shows good
agreement with a power law of $-1/5$. 
The persistence dependence represents a sensible method to
discriminate between competing models of viscoelasticity. Since the
value of the concentration scaling exponent $4/3$ predicted by an
effective medium approach \cite{morse01} is numerically quite close to
the exponent $7/5$ predicted by most other theories, experimental
accuracy does not allow for a verification. The difference of the two
concepts in the persistence length scaling exponents is considerably
larger: $-1/3$ versus $-1/5$. Contrary to recent experiments
\cite{tassieri08} our data is clearly incompatible with an exponent oft
$-1/5$. We therefore conclude that the plateau modulus of entangled
networks of semiflexible polymers is correctly described by
\begin{equation}
G \propto c^{7/5} l_p^{-1/5} \;.
\end{equation}

\section{Conclusion} \label{sec:summary}

We have presented numerical results for the plateau modulus of
entangled network of semiflexible polymers for a wide parameter range
for both an affine and non-affine shear deformation field. To this
end we have developed an approach that permits to approximately
calculate the partition sum and the free energy of a polymer network.
This was achieved by analyzing the free energy of one component of the
transversal fluctuations of a test polymer in a two dimensional
reference frame. Averaging over disorder and all possible reference
frames results in a measure for the systems free energy. The approach
allows to probe the system's free energy change and thereby mechanical
response to macroscopic and microscopic deformation fields.  While
existing theories for the modulus of polymer networks are based on the
assumption of an affine shear deformation on all length scales, we
challenged this assumptions. Indeed, it was found that the free energy
of the affine deformation can be reduced by allowing the tube contour
of a test polymer to minimize the global energy. We observed
non-affine moduli that agree well with experimental data, while
existing theoretic predictions coincide with the results of our
considerably higher affine moduli. Furthermore, we clearly confirm a
scaling of the plateau modulus with persistence length and
concentration as $G \propto c^{7/5} l_p^{-1/5}$.  Our results prove
that shear deformation of networks of entangled polymers has to be
described in a non-affine picture and that affine theories
systematically overestimate the mechanical response. The presented
approach provides a numerical solution to evaluate complex partion
sums and has a wide applicability to rheology of polymer networks.
Future applications can e.g. investigation of non-linear shear.

\begin{acknowledgement}
  We acknowledge support from the DFG through grant Fr 850/6-1, from
  the German Excellence Initiative via the NIM program and from the
  Elite Network of Bavaria through the NBT program.
\end{acknowledgement}  

\appendix
\section{Shear Deformation} \label{sec:appendix1}

\begin{figure}[htbp]
  \center
  \includegraphics[width=\columnwidth]{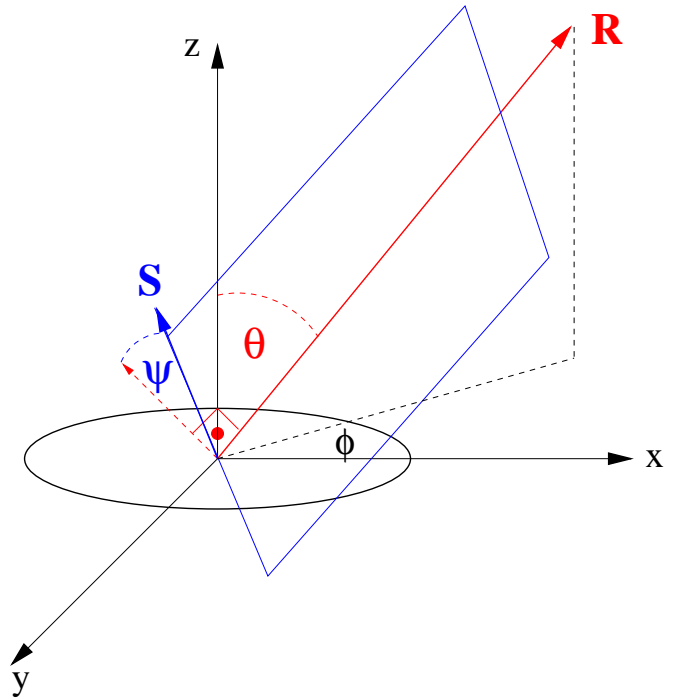}
\caption{Schematic illustration of the plane of observation.}
\label{fig:plane}
\end{figure}

We work with two coordinate systems: a three dimensional real space
system and a two dimensional system in the plane of observation, where
the origin of both systems is one end of the initial test polymer. The
orientation of the end-to-end vector ${\bf R}$ of an arbitrary test
polymer in the three dimensional space is isotropically distributed.
As depicted in Fig. \ref{fig:plane} it is described by the two angles
$\theta$ and $\phi$. To define a plane of fluctuations we need one
additional angle $\psi$. This plane is spanned by the vectors ${\bf
  R}$ and ${\bf S}$ with ${\bf S} \perp {\bf R}$. For $\psi=0$ the
vector ${\bf S}$ is obtained by applying the same transformation to an
vector parallel to the z-axis, that is needed to transform an vector
parallel to the x-axis to ${\bf R}$. Other values of $\psi$ are
obtained by an rotation around the axis ${\bf R}$. Graphically it is
helpful to picture the observation plane as the plane that is obtained
by applying two transformations to the x-z-plane: first a rotation
around the z-axis by $\phi$ and then a rotation around the axis ${\bf
  R}$ by $\psi$. If ${\bf R}$ and ${\bf S}$ are normalized they
correspond to the x and y-axis in the plane of observation.

If we now apply a macroscopic shear deformation $\mathcal{T}(\Gamma)$
with shear parameter $\Gamma$ to the three dimensional system the
obstacle points and the test polymer deform according to the action of
the transformation $\mathcal{T}$ on their real space coordinates. In
general this signifies that they leave the plane of observation
spanned by ${\bf R}$ and ${\bf S}$. It is however, self-evident that
also the test polymer's fluctuations are subjected to the shear and
therefore also the plane of observation transforms according to
$\mathcal{T}$. This is tantamount to a transformation of the vectors
${\bf R}$ and ${\bf S}$ and leaves the transformed obstacles points in
the new plane of observation. We obtain the new plane coordinates in
terms of the transformed unit vectors ${\bf R'}$ and ${\bf S'}$ that
correspond again to x and y-axis. The former is obtained as ${\bf
  R'}=\mathcal{T}(\Gamma){\bf R}$ and the latter is constructed as the
component of $\mathcal{T}(\Gamma){\bf S}$ that is orthogonal to ${\bf
  R'}$.

\bibliographystyle{unsrt}

\end{document}